\documentclass[reprint,prb,twocolumn,showpacs,superscriptaddress,aps,longbibliography,floatfix]{revtex4-1}

\usepackage[utf8]{inputenc}
\usepackage[T1]{fontenc}

\usepackage{graphicx}
\DeclareGraphicsExtensions{.png,.jpg,.eps,.pdf}

\usepackage{float}
\usepackage{xcolor}
\usepackage{amsmath}
\usepackage[colorlinks=true,citecolor=blue]{hyperref}

\begin{document}

\title{Moir\'e Mott correlated mosaics in twisted bilayer 1T-TaS$_2$}

\author{Ana Vera Montoto}
\email{Email: ana.veramontoto@aalto.fi}
\affiliation{Department of Applied Physics, Aalto University, 02150 Espoo, Finland}

\author{Jose L. Lado}
\email{Email: jose.lado@aalto.fi}
\affiliation{Department of Applied Physics, Aalto University, 02150 Espoo, Finland}

\author{Adolfo O. Fumega}
\email{Email: adolfo.oterofumega@aalto.fi}
\affiliation{Department of Applied Physics, Aalto University, 02150 Espoo, Finland}

\begin{abstract}
\textbf{ABSTRACT:} The tunability and twist engineering of van der Waals materials enable the emergence of electronic states not present in individual monolayers.
Among them, monolayer 1T-TaS$_2$ is a well-known Mott insulating system, whose star-of-David charge density wave reconstruction realizes
an emergent triangular lattice of local magnetic moments. 
Interestingly, in its bulk form, the insulating gap is not correlation-driven,
but stems from interlayer coupling.
Here, we exploit the stacking-dependent nature of the insulating gap to show that in twisted 1T-TaS$_2$ bilayers, the spatially dependent competition between many-body and single-particle gaps creates
Mott--trivial mosaic superlattices, featuring regions with local magnetic moments
and non-magnetic insulating regions.
We further demonstrate the tunability of the mosaic correlated state
with an interlayer bias, giving rise to controllable charge transfer
and quenching of correlations.
Our results establish twisted 1T-TaS$_2$ as a flexible platform to engineer
mixed spatially modulated correlated insulating phases, arising from the moir\'e profile.

\textbf{KEYWORDS:} \textit{van der Waals materials, transition metal dichalcogenides, charge density wave, moir\'e, Mott insulator}
\end{abstract}

\maketitle

Twisted van der Waals heterostructures have risen as a widely flexible platform to engineer new forms of quantum matter \cite{Andrei2021}.
The twist degree of freedom produces new superlattices, whose emergent physics arises from the spatially dependent coupling between layers.
This strategy has been demonstrated using two-dimensional semimetals \cite{Cao2018,Lu2019,Park2026}, 
semiconductors \cite{Sutter2019, PhysRevB.108.075416}, 
superconductors \cite{MARTINI2023106,Xia2024},
insulators \cite{Yasuda2021} and 
even Mott insulators \cite{Song2021,Vao2021}.
In the case of halide Mott insulators, such as CrI$_3$ \cite{Song2021}, CrBr$_3$,
CrCl$_3$, VCl$_3$ or NiI$_2$ magnets, 
the twist degree of freedom drives the system to different
magnetic states depending on the angle. However, their Mott
nature remains unaffected due to the large mismatch between
the Mott energy scale (3--4 eV) and the interlayer energy scale (100--300 meV) \cite{Fumega2023,Anto2024,Yao2023,PhysRevLett.133.246703,Akram2024,Yao2024}.

A promising family of two-dimensional Mott insulators
are 1T transition metal dichalcogenides (TMDs) \cite{Chen2020,Ruan2021,Zhang2024,PhysRevLett.134.046504}. These
include 1T-TaS$_2$, 1T-TaSe$_2$,
1T-NbSe$_2$ and 1T-NbS$_2$, which feature
a charge density wave distortion (CDW), leading to
a ``star-of-David'' (SoD) unit cell with 13 Ta (Nb) atoms and
a nearly flat band
emerging from the distortion itself \cite{PhysRevB.97.045133}.
The nearly flat band inherits the correlations from the Ta (Nb) atoms,
leading to a Mott insulating state \cite{Chen2020,Ruan2021,2025arXiv251103311W}, 
with a reduced Mott scale
of around 200 meV due to the CDW nature of the flat band \cite{PhysRevB.105.L081106}.
In the monolayer limit, this gives rise to a Mott insulating state with geometrically frustrated magnetism, becoming a potential
platform for quantum spin liquid states \cite{Ruan2021,2025arXiv251103311W}.
In contrast, 
the coupling between layers in bulk 1T-TaS$_2$ overcomes the Mott
gap and yields an insulating state that is driven by interlayer coupling
rather than Mott correlations \cite{PhysRevLett.129.016402}.
While the physics of both monolayer and bulk
have been studied in detail, potential emergent phases in twisted 1T-TaS$_2$ and its related 1T family of dichalcogenides have remained largely unexplored.
Twisted 1T-TaS$_2$ bilayers have been synthesized and their structure studied \cite{D0NR05148A}, and \textit{ab initio} methods have predicted a metal-semiconductor transition and an insulating gap opening respectively in two relaxed structures of 1T-TaS$_2$ bilayers for large twist angles \cite{LI20192302}.
But only recently have the moir\'e modulations of the electronic structure of 1T-TaSe$_2$ been probed with scanning tunneling microscopy (STM) \cite{doi:10.1073/pnas.2520703123}.
Twisted TMDs, including other such as CDW 1T-NbSe$_2$ \cite{dai_layer_2023}, 2H-TaS$_2$ \cite{hu_engineering_2022, kovalska_photocatalytic_2021} and twisted TMD heterostructures \cite{huang_angle-dependent_2024, shah_progress_2024, yuan_probing_2023, weston_atomic_2020} provide a variety of emergent phenomena achieved in recent years through twist engineering.

Here, we present a model of twisted 1T-TaS$_2$ bilayers and analyze their electronic structure.
We show that due to the spatially dependent coupling between layers, the moir\'e pattern directly imprints a profile of many-body and single-particle gaps in real space.
In particular, the magnetization profiles obtained by the model reveal that certain regions of the moir\'e develop local magnetic moments stemming from correlations, whereas others remain non-magnetic due to an interlayer-driven single-particle gap.
We also show the moir\'e modulated spectral signatures emerging from the coexisting many-body and single-particle correlated states.
Finally, we demonstrate that an interlayer bias can be used to induce spatially modulated charge transfer, leading
to a controllable local Mottness of the mosaic.
Our results establish a new van der Waals platform to engineer emergent electronic states by controlling the degree of correlations through a moir\'e pattern.

\begin{figure*}[t!]
\centering
\includegraphics[width=\linewidth]{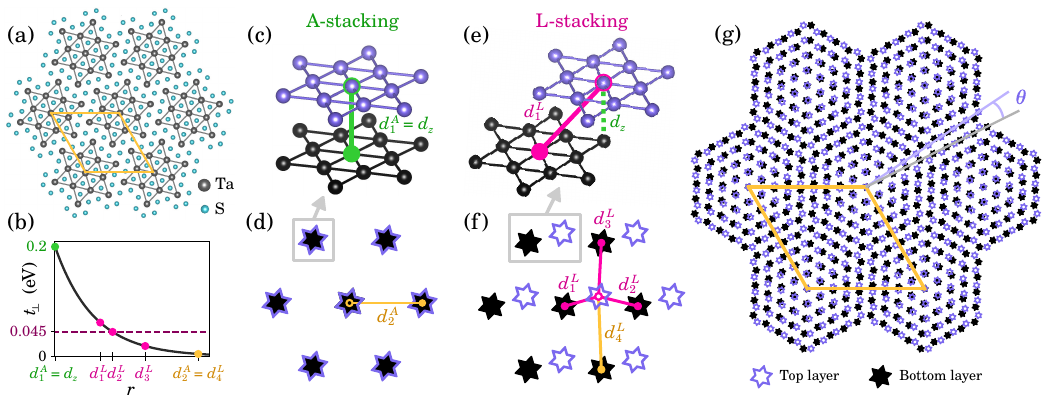}
\caption{
\textbf{(a)} CCDW atomic structure of monolayer 1T-TaS$_{2}$ (top view), depicting the closest bonds between Ta atoms. CCDW supercell delineated in yellow.
\textbf{(b)} Hopping $t_\perp$ considered in the bilayer model between two SoD sites in different layers as a function of the distance between them $r$ for $\tau = 0.2$ eV \cite{PhysRevLett.129.016402}.
\textbf{(c--f)} Stacking orders found in bulk 1T-TaS$_2$.
(c,e) Ta atoms of the 2 SoD structures that make up the unit cells of A-stacking (c) and L-stacking (e).
(d,f) Top view of the SoD sites of the A-stacked (d) and L-stacked (f) bilayers. Star symbols represent the center of each SoD structure, where the electrons are localized.
\textbf{(g)} Moir\'e structure of twisted bilayer 1T-TaS$_{2}$ with twist angle $\theta = 7.34^\circ$ (top view).
}
\label{fig:1}
\end{figure*}

1T tantalum dichalcogenides 1T-TaS$_{2}$ and 1T-TaSe$_{2}$ have a layered triangular lattice structure, and present low-temperature commensurate CDW (CCDW) phases in which groups of 13 Ta atoms are clustered to form a 6-pointed star unit cell (SoD), each hosting one conduction electron in a $d_{z^2}$ orbital localized on the central Tantalum atom \cite{rossnagel_spin-orbit_2006, nicholson_gap_2024, park_stacking_2023}. A $\sqrt{13} \times \sqrt{13}$ SoD superlattice emerges, as depicted in Fig. \ref{fig:1}(a). The electronic structure of a monolayer in the CCDW phase can be 
described with a Wannier Hamiltonian for the nearly flat bands
in a triangular lattice, in which each site represents the Wannier state
of a SoD \cite{Kim2023, PhysRevLett.129.016402, Dalal2025, Chen_2022}. We can then build a half-filled Hubbard model Hamiltonian of this triangular lattice $H_{\text{mono}} = H_{\text{intra}} + H_{U}$ to characterize the monolayer Mott insulating phase \cite{PhysRevLett.129.016402}. $H_{\text{intra}}$ is the tight-binding term that considers up to third-nearest-neighbor interactions, and the local interaction term $H_U$ is the projected onsite electronic repulsion in the Wannier orbitals.

\begin{equation}
H_{\text{intra}} = \sum_{i,j,s} t_{ij} c^{\dagger}_{i, s} c_{j, s}+ \text{h.c.},
\end{equation}

\begin{equation}
H_{U} = U \sum_{n} c^{\dagger}_{n,\uparrow} c_{n,\uparrow} c^{\dagger}_{n,\downarrow} c_{n,\downarrow},
\end{equation}

where $t_{ij}$ denotes the hopping between sites $i$ and $j$, and $c^{\dagger}_{i, s}$, $c_{i, s}$ are the creation and annihilation operators of an electron with spin polarization $s = \, \uparrow, \downarrow$ at site $i$.

The nearest-neighbor intralayer hopping parameter of 1T-TaS$_2$ is taken as $t_{1} = 0.05\,U$
\footnote{\textit{Ab initio} methods could provide a more accurate nearest-neighbour intralayer hopping (as done in the literature for 1T-TaSe$_2$ \cite{doi:10.1073/pnas.2520703123}), but the model predicts the Mott mosaic phenomenology to hold in the regime of dominant $U$ and interlayer hopping. We have included in the Supporting Information some results reproducing the Mott mosaic for the lower intralayer hopping parameters for 1T-TaSe$_2$.}
, and the second- and third-nearest-neighbor hoppings preserve the ratios $t_{2} = -0.23\,t_1$ and $t_{3} = -0.27\, t_{1}$ 
stemming from the density functional theory electronic dispersion \cite{Chen_2022}.

The interaction term $H_U$ can be solved at the mean-field level $H_{U} \approx H_{U}^{MF}$ by performing a 
non-collinear mean-field decoupling of the form $c^\dagger_{n,\uparrow}c_{n,\uparrow} c^\dagger_{n,\downarrow}c_{n,\downarrow} \approx \langle c^\dagger_{n,\uparrow}c_{n,\uparrow} \rangle c^\dagger_{n,\downarrow}c_{n,\downarrow}  +
\langle c^\dagger_{n,\downarrow}c_{n,\downarrow} \rangle c^\dagger_{n,\uparrow}c_{n,\uparrow} -
\langle c^\dagger_{n,\downarrow}c_{n,\uparrow} \rangle c^\dagger_{n,\uparrow}c_{n,\downarrow} -
\langle c^\dagger_{n,\uparrow}c_{n,\downarrow} \rangle c^\dagger_{n,\downarrow}c_{n,\uparrow}$. The previous replacement gives rise to a mean-field ansatz that is explicitly
rotationally symmetric, and enables the emergence of general non-coplanar magnetic textures.
The mean-field Hamiltonian takes the form

\begin{equation}
H_U^{\text{MF}} =
\sum_{n,\alpha,s,s'} J^\alpha_n \cdot \sigma^\alpha_{s,s'} c^\dagger_{n,s} c_{n,s'},
\end{equation}
where $\sigma^\alpha$ are the spin Pauli matrices, and $J_n^\alpha$ is the self-consistent
interaction-induced
exchange field given by $J^\alpha_n = -U/2 \sum_{s,s'}\sigma^\alpha_{s,s'}\langle c^\dagger_{n,s} c_{n,s'} \rangle $.
The interacting problem is solved by using a variational Hartree-Fock
wavefunction $| \Omega_{\text{MF}}\rangle = \prod_n \psi^\dagger_n |0 \rangle$,
where $|0 \rangle$ is the empty state and $\psi^\dagger_n$ are variational single-particle states, which are computed by solving the mean-field equations self-consistently\cite{pyqula}.

The lowest energy stacking structure of bulk 1T-TaS$_2$ (AL-stacking) exhibits the two alternating stacking arrangements displayed in Fig. \ref{fig:1}(c--f), labeled A-stacking and L-stacking in reference to the center Ta atom of the SoD (A), and its upper-right star point (L) \cite{PhysRevLett.122.106404}.
Our model will capture not only both of these stacking arrangements, but also any translation and twist angle between layers, by taking an interlayer hopping of the form

\begin{equation}
H_{\text{inter}} = \sum_{\langle ij \rangle_{\perp},\, s}
t_\perp(r_{ij}) \, c^{\dagger}_{i, s} c_{j, s} + \text{h.c.},
\end{equation}

where $\langle ij \rangle_{\perp}$ denotes a pair of sites belonging to different layers. The interlayer hopping is taken as a function that decays exponentially with the distance between two SoD sites, as shown in Fig. \ref{fig:1}(b).
\begin{equation}
t_{\perp}(r) = \tau e^{\lambda(d_{z} - r)}.
\end{equation}
$d_{z} = 5.9$ \AA{} \cite{10.1063/1.4805003} is the minimal distance between two layers of 1T-TaS$_2$.
The decay parameter $\lambda = 0.49$ \AA{}$^{-1}$ is 
chosen to reproduce bulk 1T-TaS$_{2}$ parameters fitted to STM measurements \cite{PhysRevLett.129.016402}, specifically those of the A-stacked nearest-neighbor distance $d_{1}^{A} = d_z$ ($t_{\perp}(d_{1}^{A}) = 0.2$ eV \cite{PhysRevLett.129.016402}) and the L-stacked second-nearest-neighbor distance $d_{2}^L$ ($t_{\perp}(d_{1,2,3}^{L}) = 0.045$ eV \cite{PhysRevLett.129.016402}). The value of the hopping strength factor $\tau \equiv t_{\perp}(d_{z})$ that would correspond to the aforementioned model is $\tau = 0.5 \,U$ (as $U = 0.4$ eV \cite{PhysRevLett.129.016402}). 
It is worth noting that $\tau$ can be experimentally increased with external uniaxial pressure, so in the following we
will explore the effect of this factor, while examining in greater detail a representative case in the realistic regime ($\tau = 0.7\,U$). The distance-dependent interlayer hopping term allows the model to capture generic displaced stackings, as well as twisted layer stackings in which moir\'e periodicity emerges, as illustrated in Fig. \ref{fig:1}(g). It should also be noted that the interlayer distance $d_z$ in twisted 1T-TaS$_2$ bilayers is known to be somewhat sensitive to the twist angle \cite{D0NR05148A, LI20192302} due to atomic relaxations. An overall shift could be taken into account by the model simply by changing the interlayer hopping function, having a similar effect to a change in $\tau$. A variation of the relaxed distances across the moir\'e would be another possibility that could moderately influence the mosaic phases presented.

The full Hamiltonian for a generic twisted bilayer finally becomes

\begin{equation}
H = H_{\text{intra}} +H_{\text{inter}} + H^{\text{MF}}_{U}.
\end{equation}

It is worth noting that the previous mean-field ansatz will lead to
a symmetry-broken magnetic state in the Mott regime. As a result,
potential quantum spin liquid states \cite{Yan2011,2025arXiv251103311W} are not captured with the previous formalism,
nor does it capture finite-temperature effects leading to a paramagnetic Mott state.
The mean-field ansatz nevertheless will capture the competition between Mott
physics and hybridization gap even in those regimes, as such
phenomenology stems from the competition between kinetic energy and
interactions.

\begin{figure}[t!]
\centering
\includegraphics[width=\columnwidth]{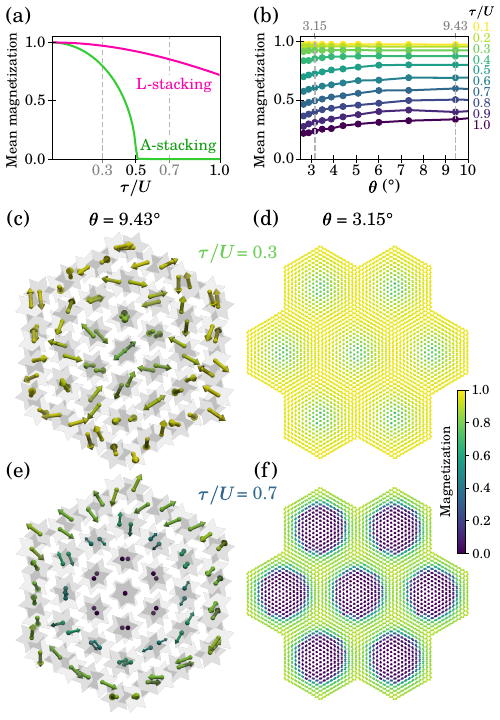}
\caption{
\textbf{(a)} Average absolute magnetization of each SoD site for the A-stacked and L-stacked 1T-TaS$_2$ bilayers. Dashed lines highlight the values of $\tau$ of the twisted calculations shown in (c--f).
\textbf{(b)} Average absolute magnetization of each SoD site for 1T-TaS$_2$ bilayers twisted at different angles $\theta$ and considering different interlayer hopping strengths $\tau$. Dashed lines highlight the angles $\theta$ shown in (c--f).
\textbf{(c--f)} Top view of the magnetization on each SoD site in 1T-TaS$_2$ bilayers with twist angles $\theta=9.43^\circ$ (c,e) and $\theta=3.15^\circ$ (d,f) in the low interlayer coupling regime, $\tau = 0.3 \,U$ (c--d), and the high coupling regime $\tau = 0.7 \,U$ (e--f). (c,e) Magnetization vectors on each SoD site of a moir\'e supercell. (d,f) Norm of the magnetization of each site for multiple moir\'e supercells.
}
\label{fig:2}
\end{figure}

We first analyze the phase diagram for the aligned bilayers as given by the model above, which can
be used as a reference for calculations of twisted bilayers. Fig. \ref{fig:2}(a) shows that the mean-field calculations using the described model predict that when increasing the interlayer hopping strength $\tau$, the magnetization of the sites in the A-stacked bilayer quickly decreases, becoming completely non-magnetic for $\tau \ge 0.5\,U$. In contrast, the magnetization of the L-stacked bilayer sites only shows a slight decrease from the monolayer case, $\tau=0$ (non-interacting layers), which is known to be a Mott insulator. The results in the regime of sizable coupling between layers, $\tau \ge 0.5\,U$, are consistent with previous predictions \cite{PhysRevLett.129.016402}, which show that a bulk sample of TaS$_2$ (A-L alternating stacking) terminating in an A-stacked bilayer opens a hybridization gap and is band insulating, while one with an L-stacked bilayer termination hosts surface Mott insulating states. In the following analysis, we consider the representative case $\tau = 0.7\,U$ in order to focus on the realistic regime.

\begin{figure*}[t!]
\centering
\includegraphics[width=\linewidth]{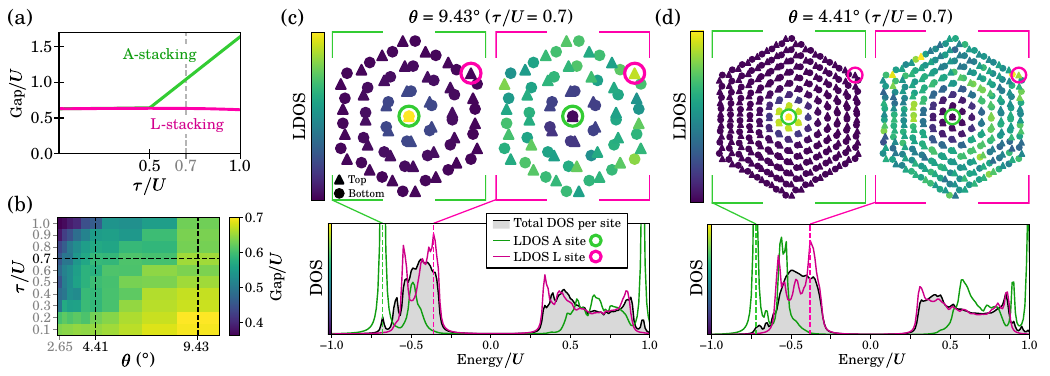}
\caption{
 \textbf{(a)} Band gap of the non-twisted A-stacked and L-stacked 1T-TaS$_2$ bilayers for different interlayer hopping strengths $\tau$. The dashed line indicates the value of $\tau$ used in plots (c--d).
 \textbf{(b)} Band gap of twisted bilayers with different twist angles $\theta$ and interlayer hopping strengths $\tau$. The value of $\tau$ and the twist angles used in plots (c--d) are highlighted.
 \textbf{(c--d)} Density of states of the 1T-TaS$_2$ bilayer with interlayer hopping strength $\tau = 0.7 \,U$ and twist angles $\theta = 9.43^\circ$ (c) and $\theta = 4.41^\circ$ (d). Bottom: Total DOS divided by the number of SoD sites in the moir\'e supercell alongside LDOS of sites A and L. Top: LDOS of every SoD site plotted as a top view in real space, at peaks of the LDOS spectrum of the A site (left) and L site (right). Sites labeled ``A'' and ``L'' are circled.
}
\label{fig:3}
\end{figure*}

We now address moir\'e bilayers at different twist angles. Adding a twist angle between the two layers introduces local variations of interlayer neighbor distances, which in turn modulate the local
interlayer hopping. We observe (Fig. \ref{fig:2}) that sites with closer interlayer neighbors (A-like regions) have reduced magnetization compared with sites whose interlayer neighbors are farther away (L‑like regions). The effect is particularly strong in the cases where  $\tau > 0.5\,U$, as seen in Fig. \ref{fig:2}(e--f) for $\tau =0.7\,U$ with twist angles $\theta = 9.43^\circ$ and $\theta = 3.15^\circ$, respectively. In these cases, non-magnetic regions appear in the center of the moir\'e supercells, where the interlayer coupling between SoD sites is similar to that of the A-stacking order. Even for a low interlayer hopping strength $\tau = 0.3\,U$ in Fig. \ref{fig:2}(c--d), the A--L contrast remains noticeable. This is analogous to the difference in the non-twisted results for these $\tau$ values (see dashed lines in Fig. \ref{fig:2}(a) for comparison). Fig. \ref{fig:2}(b) shows the average absolute local magnetization 
for different commensurate moir\'e twist angles and interlayer hopping strength values. As expected from the non-twisted results in Fig. \ref{fig:2}(a), Fig. \ref{fig:2}(b) demonstrates the effect of higher hopping strength between the layers decreasing the mean magnetization. Along with it, we observe that decreasing the twist angle has a smaller effect in lowering the mean magnetization.

It is worth noting that the local quenching of the Mott state can enable the promotion of
further exotic magnetic states, beyond the symmetry-broken solutions considered
in our calculations. 
In particular, superlattice effects alone are potentially able to drive
a magnetically ordered state into a quantum spin liquid state.
While a purely triangular Heisenberg model shows a magnetically ordered state,
a $2 \times 2$ triangular superlattice with a quenched site realizes a Kagome
lattice, which shows a quantum spin liquid regime \cite{Yan2011}.
A further paradigmatic example is the maple-leaf lattice \cite{2026arXiv260105308E,2025arXiv251220466N,2025arXiv251121598G},
which can be understood as a locally quenched antiferromagnetic triangular superlattice.
Recent results show that such a model sustains a variety of frustrated states, 
including spin liquids, enabled by the enlarged superlattice structure. 
These suggest that twisted bilayers can provide a potential
platform to realize quantum magnetic states even when the monolayer develops magnetic order.
Such spin liquid states, however, would require a methodology
beyond the one presented in our manuscript, requiring either
tensor network methods or approximate auxiliary fermion calculations.

Another point of note is that the inclusion of spin-orbit coupling in the model could affect the emergent magnetic phase in the moir\'e system and renormalize the gap. However, the Mott mosaic phenomenology obtained by this model is not qualitatively affected by the inclusion of spin-orbit coupling, as the interplay between interlayer hopping and onsite interactions is the driving mechanism for the Mott mosaic phases. Strong anisotropy could drive the ground state to a more collinear AFM between layers \cite{Fumega2023, Yao2023, Yao2024} as compared to the non-collinear spin texture obtained through this mean-field approach. Alternatively, the combination of spin-orbit coupling and non-collinear magnetism could also induce emergent ferroelectric polarization through the inverse Dzyaloshinskii–Moriya interaction \cite{Fumega2023, Anto2024, PhysRevLett.133.246703}.

We now analyze the spectroscopic signatures, which are directly accessible with STM \cite{2025arXiv251103311W,doi:10.1073/pnas.2520703123}. We first show how the model accounts for the spectroscopy of non-twisted bilayer stackings. Fig. \ref{fig:3}(a) shows the band gap dependence as a function of interlayer hopping strength for both A and L-stackings, indicating that at $\tau = 0.5\,U$ the hybridization gap of the A-stacked bilayer starts growing linearly with $\tau$. 
This demonstrates the band insulating character of the A-stacked bilayer in the $\tau > 0.5\,U$ regime, while the L-stacked bilayer remains a Mott insulator (see Supporting Information for an extension of the model to phases beyond $\tau>U$).

\begin{figure*}[t!]
\centering
\includegraphics[width=\linewidth]{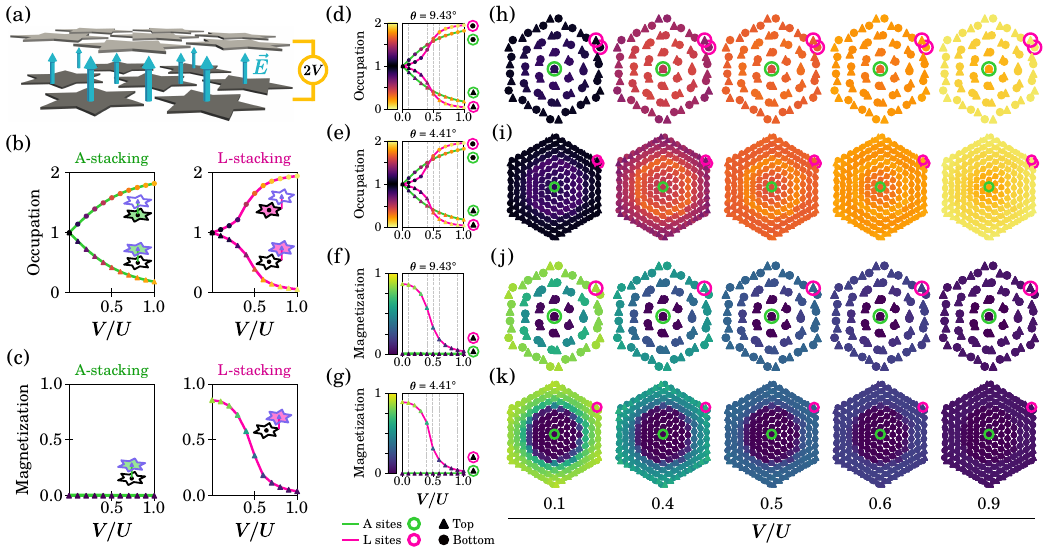}
\caption{
\textbf{(a)} Schematic of a bias voltage $2V$ applied between two twisted 1T-TaS$_2$ layers.
\textbf{(b)} Occupation of the top and bottom sites of A-stacking in contrast with those of L-stacking with increasing $V$, for interlayer hopping strength $\tau = 0.7 \, U$.
\textbf{(c)} Magnetization of the top sites of A- and L-stacking with increasing interlayer bias $V$, for $\tau = 0.7 \, U$.
\textbf{(d--e, h--i)} Occupation of the A and L sites (d--e) and of each site of the moir\'e supercell (h--i) of bilayer structures with twist angles $\theta = 9.43^\circ$ (d,h) and $\theta = 4.41^\circ$ (e,i), with increasing $V$, for $\tau = 0.7 \, U$.
\textbf{(f--g, j--k)} Magnetization of the top A and L sites (f--g) and of each site of the moir\'e supercell (j--k) of bilayer structures with twist angles $\theta = 9.43^\circ$ (f,j) and $\theta = 4.41^\circ$ (g,k), with increasing $V$, for $\tau = 0.7 \, U$. The corresponding L sites are circled (h--k). Note that the magnetization in the bottom sites, not shown in (c,f--g) is equal in magnitude with opposite direction.}
\label{fig:4}
\end{figure*}

We can then calculate the local density of states (LDOS) of the twisted bilayer structures in the realistic $\tau$ regime and compare sites with a close interlayer neighbor, labeled ``A'', against sites with farther neighbors, labeled ``L''.  Fig. \ref{fig:3}(b) shows the band gap for 1T-TaS$_2$ twisted bilayer systems and its dependence on $\tau$. The LDOS results indicate that the smallest gap corresponds to the L sites, and so it is expected that the gap decreases with higher $\tau$ values. Additionally, we observe an increase in this Mott gap for larger twist angles. As seen in the lower panels of Fig. \ref{fig:3}(c--d) for two different twist angles, the gap of the A-site LDOS is significantly larger than the L-site LDOS for any twist angle, consistent with hybridization in the central region of the moir\'e supercells where the sites are non-magnetic.

Furthermore, the difference in the LDOS gaps produces modulations of the LDOS in real space, as shown in the top real-space plots of Fig. \ref{fig:3}(c--d). Two examples of this moir\'e modulation are illustrated for each twist angle, one corresponding to a peak in the A-site LDOS (green) and another to a peak in the L-site LDOS (magenta). This spatial LDOS modulation in twisted bilayer TaS$_2$ could be experimentally resolved using STM, and in fact for twisted bilayer 1T-TaSe$_2$, with a similar CDW structure, spatial modulation of the gaps has been found with this technique \cite{doi:10.1073/pnas.2520703123}, although in this case the Mott insulating state is suppressed and becomes metallic. An extension of the model for higher interlayer coupling is included in the Supporting Information in order to better capture the behaviour of 1T-TaSe$_2$.

\begin{figure*}[t!]
\centering
\includegraphics[width=\linewidth]{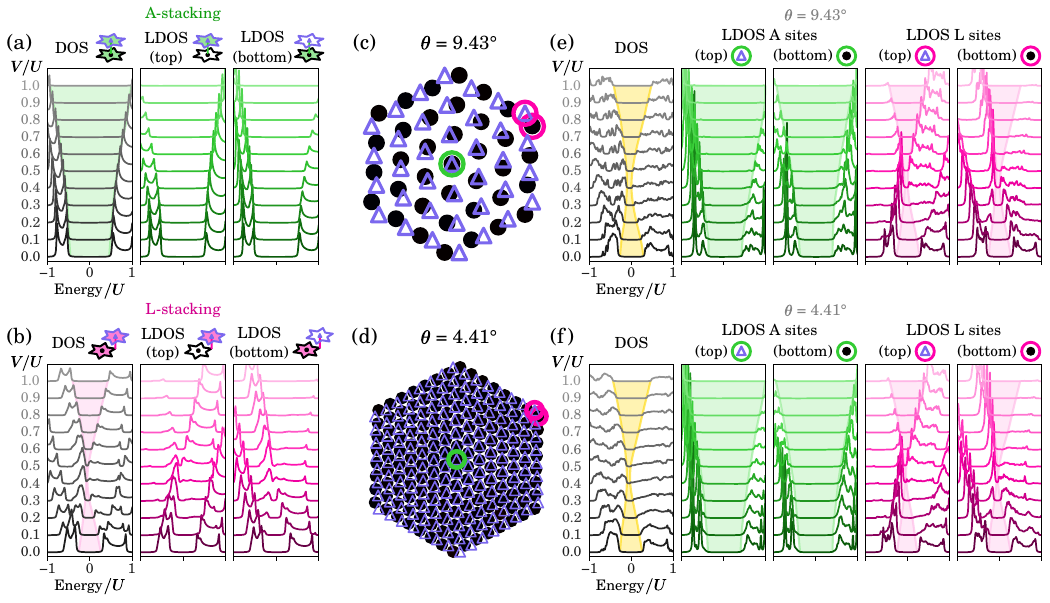}
\caption{
\textbf{(a--b)} DOS of the A-stacked (a) and L-stacked (b) 1T-TaS$_2$ bilayers with interlayer hopping strength $\tau = 0.7 \, U$ and LDOS of their top and bottom sites for increasing interlayer bias $2V$.
\textbf{(c--d)} Top view of bilayer lattice structures with twist angles  $\theta = 9.43^\circ$ (c) and $\theta = 4.41^\circ$ (d), with A sites marked by a green circle and L sites marked by magenta circles. \textbf{(e--f)} For the 1T-TaS$_2$ bilayers with $\tau = 0.7 \, U$ and twist angle $\theta = 9.43^\circ$ (e) and $\theta = 4.41^\circ$ (f): total DOS, LDOS at the A sites in the top and bottom layers, and LDOS at the L sites in the top and bottom layers.}
\label{fig:5}
\end{figure*}

The low dimensionality of twisted van der Waals bilayers enables a high degree of tunability through external electric gates \cite{Han2025,Xia2024,doi:10.1021/acs.chemrev.3c00627,Park2022,Rickhaus2018,Rickhaus2019,Xu2019,Zhang2023,Zhou2021,Park2021}. 
Therefore, having characterized the emergent moir\'e Mott mosaic in twisted 1T-TaS$_2$ bilayers, we show how the correlated state can be externally controllable with an external interlayer gate bias. 
A bias between layers gives rise to
an additional term in the Hamiltonian of the form

\begin{equation}
H_V = V \sum_{n,s} \chi_n c^\dagger_{n,s} c_{n,s}
\end{equation}

where $\chi_n=\pm 1$ for the upper and lower layer, respectively, and $V$ controls the strength of the interlayer bias ($2V$), as represented in Fig. \ref{fig:4}(a). In particular, such a term will drive the monolayers away
from half filling, promoting charge transfer from one layer to the other \cite{PhysRevB.110.195138,Crippa2024,PhysRevLett.134.046504}.
As the bias term $V$ increases, charge transfer occurs at different rates in the A- and L-stacked bilayers, as shown in Fig. \ref{fig:4}(b). At low interlayer bias, the Mott insulating L-stacked -bilayer is more resistant to charge transfer. However, at $V \approx 0.5\,U$, charge transfer occurs at an increased rate, and so for larger bias the difference in occupation between layers becomes larger for L-stacking. This pattern also arises in the twisted bilayers, as shown in Fig. \ref{fig:4}(d--e) when comparing the occupation of the A sites against the L sites. Fig. \ref{fig:4}(h--i) displays the occupation of every site of the moir\'e supercells in real space, revealing the resulting bias-dependent modulation at the moir\'e scale.
We further analyze the magnetization profiles with increasing $V$. Fig. \ref{fig:4}(c) shows that the A-stacked bilayer remains non-magnetic as expected, whereas the 
magnetization in the L-stacking sites decreases at a rate similar to the change in their occupation. 
This behavior is found locally in the A and L sites of the twisted layers, as shown in Fig. \ref{fig:4}(f--g); and Fig. \ref{fig:4}(j--k) illustrate this decreasing magnetization of the antiferromagnetic Mott regions as $V$ increases.

We now address the effect of interlayer bias on the density of states of 
the bilayer systems. In the non-twisted structures, a
distinct behavior of the two stacking orders in the presence of $V$ appears. Increasing $V$ gradually increases the size of the gap in the A-stacked bilayer, as shown in Fig. \ref{fig:5}(a). In contrast, Fig. \ref{fig:5}(b) presents a more dramatic electrical control of the electronic structure of the L-stacked bilayer, as the gap initially decreases, closing near $V \approx 0.5\,U$, and then increasing for higher bias. The LDOS of the A and L sites of the twisted bilayers, located in the center and edge of the moir\'e supercell as depicted in Fig. \ref{fig:5}(c--d), locally reflect the features observed
in the A and L-stacking orders: LDOS gap increasing in the A sites, but closing and reopening in the L sites.

This dependence of the gap on an external bias provides an additional experimental signature of the Mott region of the moir\'e. Other possibilities to probe the Mott phases would include resolving the magnetic texture through spin-polarized STM techniques and using Inelastic Tunneling spectroscopy (IETS) to measure spin excitations (magnons or spinons) associated with the magnetic phases emerging in the Mott regions of the moir\'e system, as demonstrated in multiple experiments
in 2D materials \cite{PhysRevLett.102.256802, 2025arXiv251103311W, Turco2024, Zhao2025, Mishra2021}. Additionally, Kondo resonances at zero bias could provide solid proof of the Mott character of the gap \cite{fei_understanding_2022}. The spin response of the twisted bilayer system that could be probed with IETS is included in the Supporting Information.

The bias tunability of the Mott region (L sites), in contrast
with the minimal changes in the A sites, may also indicate that changes in the electronic structure of the twisted bilayer system resulting from doping or superconducting states would be localized to this Mott region.
These results establish the potential for electrically fine-tunable semiconducting devices based on the moir\'e Mott mosaic emerging in twisted 1T-TaS$_2$ bilayers.

We have shown that the spatially dependent coupling
in 1T-TaS$_2$ leads to a competition between single-particle and correlation-driven Mott
gaps. The moir\'e bilayer features a fully gapped state across the whole moir\'e,
yet the origin of the gap is directly inherited from the local stacking of the moir\'e pattern.
We have shown the emergence of a moir\'e mosaic with an arrangement of magnetic and non-magnetic regions, corresponding to Mott and hybridized gaps, respectively.
We demonstrated that the moir\'e Mott mosaic is tunable with an electric bias in a non-monotonic way thanks to the coexistence of Mott and single-particle regions.
Our results establish 1T-TaS$_2$ twisted bilayers as promising
candidates for tunable electronic and magnetic phases, providing a unique platform
where Mott and single-particle gaps are directly imprinted through twist engineering.

\smallskip
\textbf{Supporting Information}:
Discussion of the generality of the model and extended calculations to lower intralayer and higher interlayer hopping as predicted for 1T-TaSe$_2$. Simulation of IETS signatures of the Mott mosaic.

\smallskip
\textbf{Acknowledgements}:
We acknowledge the computational resources provided by the Aalto Science-IT project and the financial support from InstituteQ, the Finnish Quantum Flagship, the Finnish Ministry of Education and Culture through the Quantum Doctoral Education Pilot Program (QDOC VN/3137/2024-OKM-4), the Research Council of Finland (no. 370912, 369367 and 358877), the Finnish Centre of Excellence in Quantum Materials QMAT (no. 374166), and
the ERC Consolidator Grant ULTRATWISTRONICS (Grant agreement no. 101170477).
We thank Z. Wang, R. Drost and P. Liljeroth for useful discussions.

\bibliography{main.bib}

@article{Andrei2021,
  title = {The marvels of moir\'e materials},
  volume = {6},
  ISSN = {2058-8437},
  url = {http://dx.doi.org/10.1038/s41578-021-00284-1},
  DOI = {10.1038/s41578-021-00284-1},
  number = {3},
  journal = {Nature Reviews Materials},
  publisher = {Springer Science and Business Media LLC},
  author = {Andrei,  Eva Y. and Efetov,  Dmitri K. and Jarillo-Herrero,  Pablo and MacDonald,  Allan H. and Mak,  Kin Fai and Senthil,  T. and Tutuc,  Emanuel and Yazdani,  Ali and Young,  Andrea F.},
  year = {2021},
  month = mar,
  pages = {201–206}
}

@ARTICLE{2026arXiv260105308E,
       author = {{Ebert}, Paul L. and {Iqbal}, Yasir and {Wietek}, Alexander},
        title = {Competing Paramagnetic Phases in the Maple-Leaf {Heisenberg} Antiferromagnet},
      journal = {arXiv e-prints},
         year = 2026,
        month = jan,
          eid = {arXiv:2601.05308},
        pages = {arXiv:2601.05308},
          doi = {10.48550/arXiv.2601.05308},
 primaryClass = {cond-mat.str-el},
       adsurl = {https://ui.adsabs.harvard.edu/abs/2026arXiv260105308E},
      note    = {(accessed 2026-08-16)}
}

@article{Yan2011,
  title = {Spin-Liquid Ground State of the {S} = 1/2 Kagome {Heisenberg} Antiferromagnet},
  volume = {332},
  ISSN = {1095-9203},
  url = {http://dx.doi.org/10.1126/science.1201080},
  DOI = {10.1126/science.1201080},
  number = {6034},
  journal = {Science},
  publisher = {American Association for the Advancement of Science (AAAS)},
  author = {Yan,  Simeng and Huse,  David A. and White,  Steven R.},
  year = {2011},
  month = jun,
  pages = {1173–1176}
}

@ARTICLE{2025arXiv251121598G,
       author = {{Gresista}, Lasse and {Kiese}, Dominik and {Trebst}, Simon and {Iqbal}, Yasir},
        title = "{Unconventional orders in the maple-leaf ferro-antiferromagnetic {Heisenberg} model}",
      journal = {arXiv e-prints},
         year = 2025,
        month = nov,
          eid = {arXiv:2511.21598},
        pages = {arXiv:2511.21598},
          doi = {10.48550/arXiv.2511.21598},
 primaryClass = {cond-mat.str-el},
       adsurl = {https://ui.adsabs.harvard.edu/abs/2025arXiv251121598G},
      note    = {(accessed 2026-08-16)}
}

@ARTICLE{2025arXiv251220466N,
       author = {{Nyckees}, Samuel and {Ghosh}, Pratyay and {Mila}, Fr{\'e}d{\'e}ric},
        title = "{Tensor-network study of the ground state of maple-leaf {Heisenberg} antiferromagnet}",
      journal = {arXiv e-prints},
         year = 2025,
        month = dec,
          eid = {arXiv:2512.20466},
        pages = {arXiv:2512.20466},
          doi = {10.48550/arXiv.2512.20466},
 primaryClass = {cond-mat.str-el},
       adsurl = {https://ui.adsabs.harvard.edu/abs/2025arXiv251220466N},
      note    = {(accessed 2026-08-16)}
}

@ARTICLE{2025arXiv251103311W,
       author = {{Wang}, Ziying and {Fumega}, Adolfo O. and {Montoto}, Ana Vera and {Amini}, Mohammad and {Gamze Arslan}, B{\"u}{\textcommabelow s}ra and {Cahl{\'\i}k}, Ale{\v{s}} and {Ding}, Yuxiao and {Lado}, Jose L. and {Drost}, Robert and {Liljeroth}, Peter},
        title = {Moir{\'e} modulated quantum spin liquid candidate {1T-TaSe$_2$}},
      journal = {arXiv e-prints},
         year = 2025,
        month = nov,
          eid = {arXiv:2511.03311},
        pages = {arXiv:2511.03311},
          doi = {10.48550/arXiv.2511.03311},
 primaryClass = {cond-mat.str-el},
       adsurl = {https://ui.adsabs.harvard.edu/abs/2025arXiv251103311W},
      note    = {(accessed 2026-08-16)}
}

@article{Yasuda2021,
  title = {Stacking-engineered ferroelectricity in bilayer boron nitride},
  volume = {372},
  ISSN = {1095-9203},
  url = {http://dx.doi.org/10.1126/science.abd3230},
  DOI = {10.1126/science.abd3230},
  number = {6549},
  journal = {Science},
  publisher = {American Association for the Advancement of Science (AAAS)},
  author = {Yasuda,  Kenji and Wang,  Xirui and Watanabe,  Kenji and Taniguchi,  Takashi and Jarillo-Herrero,  Pablo},
  year = {2021},
  month = jun,
  pages = {1458–1462}
}

@article{Lu2019,
  title = {Superconductors,  orbital magnets and correlated states in magic-angle bilayer graphene},
  volume = {574},
  ISSN = {1476-4687},
  url = {http://dx.doi.org/10.1038/s41586-019-1695-0},
  DOI = {10.1038/s41586-019-1695-0},
  number = {7780},
  journal = {Nature},
  publisher = {Springer Science and Business Media LLC},
  author = {Lu,  Xiaobo and Stepanov,  Petr and Yang,  Wei and Xie,  Ming and Aamir,  Mohammed Ali and Das,  Ipsita and Urgell,  Carles and Watanabe,  Kenji and Taniguchi,  Takashi and Zhang,  Guangyu and Bachtold,  Adrian and MacDonald,  Allan H. and Efetov,  Dmitri K.},
  year = {2019},
  month = oct,
  pages = {653–657}
}

@article{Rickhaus2019,
  title = {Gap Opening in Twisted Double Bilayer Graphene by Crystal Fields},
  volume = {19},
  ISSN = {1530-6992},
  url = {http://dx.doi.org/10.1021/acs.nanolett.9b03660},
  DOI = {10.1021/acs.nanolett.9b03660},
  number = {12},
  journal = {Nano Letters},
  publisher = {American Chemical Society (ACS)},
  author = {Rickhaus,  Peter and Zheng,  Giulia and Lado,  Jose L. and Lee,  Yongjin and Kurzmann,  Annika and Eich,  Marius and Pisoni,  Riccardo and Tong,  Chuyao and Garreis,  Rebekka and Gold,  Carolin and Masseroni,  Michele and Taniguchi,  Takashi and Wantanabe,  Kenji and Ihn,  Thomas and Ensslin,  Klaus},
  year = {2019},
  month = oct,
  pages = {8821–8828}
}

@article{Xu2019,
  title = {Giant oscillations in a triangular network of one-dimensional states in marginally twisted graphene},
  volume = {10},
  ISSN = {2041-1723},
  url = {http://dx.doi.org/10.1038/s41467-019-11971-7},
  DOI = {10.1038/s41467-019-11971-7},
  number = {1},
  journal = {Nature Communications},
  publisher = {Springer Science and Business Media LLC},
  author = {Xu,  S. G. and Berdyugin,  A. I. and Kumaravadivel,  P. and Guinea,  F. and Krishna Kumar,  R. and Bandurin,  D. A. and Morozov,  S. V. and Kuang,  W. and Tsim,  B. and Liu,  S. and Edgar,  J. H. and Grigorieva,  I. V. and Fal’ko,  V. I. and Kim,  M. and Geim,  A. K.},
  year = {2019},
  month = sep 
}

@article{Zhang2023,
  title = {Enhanced superconductivity in spin–orbit proximitized bilayer graphene},
  volume = {613},
  ISSN = {1476-4687},
  url = {http://dx.doi.org/10.1038/s41586-022-05446-x},
  DOI = {10.1038/s41586-022-05446-x},
  number = {7943},
  journal = {Nature},
  publisher = {Springer Science and Business Media LLC},
  author = {Zhang,  Yiran and Polski,  Robert and Thomson,  Alex and Lantagne-Hurtubise,  Étienne and Lewandowski,  Cyprian and Zhou,  Haoxin and Watanabe,  Kenji and Taniguchi,  Takashi and Alicea,  Jason and Nadj-Perge,  Stevan},
  year = {2023},
  month = jan,
  pages = {268–273}
}

@article{Zhou2021,
  title = {Superconductivity in rhombohedral trilayer graphene},
  volume = {598},
  ISSN = {1476-4687},
  url = {http://dx.doi.org/10.1038/s41586-021-03926-0},
  DOI = {10.1038/s41586-021-03926-0},
  number = {7881},
  journal = {Nature},
  publisher = {Springer Science and Business Media LLC},
  author = {Zhou,  Haoxin and Xie,  Tian and Taniguchi,  Takashi and Watanabe,  Kenji and Young,  Andrea F.},
  year = {2021},
  month = sep,
  pages = {434–438}
}

@article{Park2022,
  title = {Robust superconductivity in magic-angle multilayer graphene family},
  volume = {21},
  ISSN = {1476-4660},
  url = {http://dx.doi.org/10.1038/s41563-022-01287-1},
  DOI = {10.1038/s41563-022-01287-1},
  number = {8},
  journal = {Nature Materials},
  publisher = {Springer Science and Business Media LLC},
  author = {Park,  Jeong Min and Cao,  Yuan and Xia,  Li-Qiao and Sun,  Shuwen and Watanabe,  Kenji and Taniguchi,  Takashi and Jarillo-Herrero,  Pablo},
  year = {2022},
  month = jul,
  pages = {877–883}
}

@article{Rickhaus2018,
  title = {Transport Through a Network of Topological Channels in Twisted Bilayer Graphene},
  volume = {18},
  ISSN = {1530-6992},
  url = {http://dx.doi.org/10.1021/acs.nanolett.8b02387},
  DOI = {10.1021/acs.nanolett.8b02387},
  number = {11},
  journal = {Nano Letters},
  publisher = {American Chemical Society (ACS)},
  author = {Rickhaus,  Peter and Wallbank,  John and Slizovskiy,  Sergey and Pisoni,  Riccardo and Overweg,  Hiske and Lee,  Yongjin and Eich,  Marius and Liu,  Ming-Hao and Watanabe,  Kenji and Taniguchi,  Takashi and Ihn,  Thomas and Ensslin,  Klaus},
  year = {2018},
  month = oct,
  pages = {6725–6730}
}

@article{Han2025,
  title = {Signatures of chiral superconductivity in rhombohedral graphene},
  volume = {643},
  ISSN = {1476-4687},
  url = {http://dx.doi.org/10.1038/s41586-025-09169-7},
  DOI = {10.1038/s41586-025-09169-7},
  number = {8072},
  journal = {Nature},
  publisher = {Springer Science and Business Media LLC},
  author = {Han,  Tonghang and Lu,  Zhengguang and Hadjri,  Zach and Shi,  Lihan and Wu,  Zhenghan and Xu,  Wei and Yao,  Yuxuan and Cotten,  Armel A. and Sharifi Sedeh,  Omid and Weldeyesus,  Henok and Yang,  Jixiang and Seo,  Junseok and Ye,  Shenyong and Zhou,  Muyang and Liu,  Haoyang and Shi,  Gang and Hua,  Zhenqi and Watanabe,  Kenji and Taniguchi,  Takashi and Xiong,  Peng and Zumb\"{u}hl,  Dominik M. and Fu,  Liang and Ju,  Long},
  year = {2025},
  month = may,
  pages = {654–661}
}

@article{Xia2024,
  title = {Superconductivity in twisted bilayer {WSe$_2$}},
  volume = {637},
  ISSN = {1476-4687},
  url = {http://dx.doi.org/10.1038/s41586-024-08116-2},
  DOI = {10.1038/s41586-024-08116-2},
  number = {8047},
  journal = {Nature},
  publisher = {Springer Science and Business Media LLC},
  author = {Xia,  Yiyu and Han,  Zhongdong and Watanabe,  Kenji and Taniguchi,  Takashi and Shan,  Jie and Mak,  Kin Fai},
  year = {2024},
  month = oct,
  pages = {833–838}
}

@article{Park2021,
  title = {Tunable strongly coupled superconductivity in magic-angle twisted trilayer graphene},
  volume = {590},
  ISSN = {1476-4687},
  url = {http://dx.doi.org/10.1038/s41586-021-03192-0},
  DOI = {10.1038/s41586-021-03192-0},
  number = {7845},
  journal = {Nature},
  publisher = {Springer Science and Business Media LLC},
  author = {Park,  Jeong Min and Cao,  Yuan and Watanabe,  Kenji and Taniguchi,  Takashi and Jarillo-Herrero,  Pablo},
  year = {2021},
  month = feb,
  pages = {249–255}
}

@article{Cao2018,
  title = {Correlated insulator behaviour at half-filling in magic-angle graphene superlattices},
  volume = {556},
  ISSN = {1476-4687},
  url = {http://dx.doi.org/10.1038/nature26154},
  DOI = {10.1038/nature26154},
  number = {7699},
  journal = {Nature},
  publisher = {Springer Science and Business Media LLC},
  author = {Cao,  Yuan and Fatemi,  Valla and Demir,  Ahmet and Fang,  Shiang and Tomarken,  Spencer L. and Luo,  Jason Y. and Sanchez-Yamagishi,  Javier D. and Watanabe,  Kenji and Taniguchi,  Takashi and Kaxiras,  Efthimios and Ashoori,  Ray C. and Jarillo-Herrero,  Pablo},
  year = {2018},
  month = mar,
  pages = {80–84}
}

@article{Park2026,
  title = {Experimental evidence for nodal superconducting gap in moir\'e graphene},
  volume = {391},
  ISSN = {1095-9203},
  url = {http://dx.doi.org/10.1126/science.adv8376},
  DOI = {10.1126/science.adv8376},
  number = {6780},
  journal = {Science},
  publisher = {American Association for the Advancement of Science (AAAS)},
  author = {Park,  Jeong Min and Sun,  Shuwen and Watanabe,  Kenji and Taniguchi,  Takashi and Jarillo-Herrero,  Pablo},
  year = {2026},
  month = jan,
  pages = {79–83}
}

@article{Vao2021,
  title = {Artificial heavy fermions in a van der {Waals} heterostructure},
  volume = {599},
  ISSN = {1476-4687},
  url = {http://dx.doi.org/10.1038/s41586-021-04021-0},
  DOI = {10.1038/s41586-021-04021-0},
  number = {7886},
  journal = {Nature},
  publisher = {Springer Science and Business Media LLC},
  author = {Vaňo,  Viliam and Amini,  Mohammad and Ganguli,  Somesh C. and Chen,  Guangze and Lado,  Jose L. and Kezilebieke,  Shawulienu and Liljeroth,  Peter},
  year = {2021},
  month = nov,
  pages = {582–586}
}

@article{PhysRevLett.134.046504,
  title = {Doped {Mott} Phase and Charge Correlations in Monolayer {$1T\text{\ensuremath{-}}{\mathrm{NbSe}}_{2}$}},
  author = {Huang, Xin and Lado, Jose L. and Sainio, Jani and Liljeroth, Peter and Ganguli, Somesh Chandra},
  journal = {Phys. Rev. Lett.},
  volume = {134},
  issue = {4},
  pages = {046504},
  numpages = {7},
  year = {2025},
  month = {Jan},
  publisher = {American Physical Society},
  doi = {10.1103/PhysRevLett.134.046504},
  url = {https://link.aps.org/doi/10.1103/PhysRevLett.134.046504}
}

@article{PhysRevB.110.195138,
  title = {Charge transfer in heterostructures of {$T$} and {$H$} transition metal dichalcogenides},
  author = {S\'anchez-Ram\'{\i}rez, Iri\'an and Vergniory, Maia G. and de Juan, Fernando},
  journal = {Phys. Rev. B},
  volume = {110},
  issue = {19},
  pages = {195138},
  numpages = {11},
  year = {2024},
  month = {Nov},
  publisher = {American Physical Society},
  doi = {10.1103/PhysRevB.110.195138},
  url = {https://link.aps.org/doi/10.1103/PhysRevB.110.195138}
}

@article{PhysRevLett.129.016402,
  title = {Mott versus Hybridization Gap in the Low-Temperature Phase of {$1T\text{\ensuremath{-}}{\mathrm{TaS}}_{2}$}},
  author = {Petocchi, Francesco and Nicholson, Christopher W. and Salzmann, Bjoern and Pasquier, Diego and Yazyev, Oleg V. and Monney, Claude and Werner, Philipp},
  journal = {Phys. Rev. Lett.},
  volume = {129},
  issue = {1},
  pages = {016402},
  numpages = {6},
  year = {2022},
  month = {Jun},
  publisher = {American Physical Society},
  doi = {10.1103/PhysRevLett.129.016402},
  url = {https://link.aps.org/doi/10.1103/PhysRevLett.129.016402}
}

@article{Zhang2024,
  title = {Quantum spin liquid signatures in monolayer {1T-NbSe$_2$}},
  volume = {15},
  ISSN = {2041-1723},
  url = {http://dx.doi.org/10.1038/s41467-024-46612-1},
  DOI = {10.1038/s41467-024-46612-1},
  number = {1},
  journal = {Nature Communications},
  publisher = {Springer Science and Business Media LLC},
  author = {Zhang,  Quanzhen and He,  Wen-Yu and Zhang,  Yu and Chen,  Yaoyao and Jia,  Liangguang and Hou,  Yanhui and Ji,  Hongyan and Yang,  Huixia and Zhang,  Teng and Liu,  Liwei and Gao,  Hong-Jun and Jung,  Thomas A. and Wang,  Yeliang},
  year = {2024},
  month = mar 
}

@article{Crippa2024,
  title = {Heavy fermions vs doped {Mott} physics in heterogeneous {Ta}-dichalcogenide bilayers},
  volume = {15},
  ISSN = {2041-1723},
  url = {http://dx.doi.org/10.1038/s41467-024-45392-y},
  DOI = {10.1038/s41467-024-45392-y},
  number = {1},
  journal = {Nature Communications},
  publisher = {Springer Science and Business Media LLC},
  author = {Crippa,  Lorenzo and Bae,  Hyeonhu and Wunderlich,  Paul and Mazin,  Igor I. and Yan,  Binghai and Sangiovanni,  Giorgio and Wehling,  Tim and Valentí,  Roser},
  year = {2024},
  month = feb 
}

@article{PhysRevB.105.L081106,
  title = {\textit{Ab initio} theory of magnetism in two-dimensional {1$T$\text{\ensuremath{-}}TaS$_2$}},
  author = {Pasquier, Diego and Yazyev, Oleg V.},
  journal = {Phys. Rev. B},
  volume = {105},
  issue = {8},
  pages = {L081106},
  numpages = {6},
  year = {2022},
  month = {Feb},
  publisher = {American Physical Society},
  doi = {10.1103/PhysRevB.105.L081106},
  url = {https://link.aps.org/doi/10.1103/PhysRevB.105.L081106}
}

@misc{pyqula,
  author = {Jose L. Lado},
  title = {https://github.com/joselado/pyqula},
  url = {https://github.com/joselado/pyqula},
  year = {2026}
}

@article{Ruan2021,
  title = {Evidence for quantum spin liquid behaviour in single-layer {1T-TaSe$_2$} from scanning tunnelling microscopy},
  volume = {17},
  ISSN = {1745-2481},
  url = {http://dx.doi.org/10.1038/s41567-021-01321-0},
  DOI = {10.1038/s41567-021-01321-0},
  number = {10},
  journal = {Nature Physics},
  publisher = {Springer Science and Business Media LLC},
  author = {Ruan,  Wei and Chen,  Yi and Tang,  Shujie and Hwang,  Jinwoong and Tsai,  Hsin-Zon and Lee,  Ryan L. and Wu,  Meng and Ryu,  Hyejin and Kahn,  Salman and Liou,  Franklin and Jia,  Caihong and Aikawa,  Andrew and Hwang,  Choongyu and Wang,  Feng and Choi,  Yongseong and Louie,  Steven G. and Lee,  Patrick A. and Shen,  Zhi-Xun and Mo,  Sung-Kwan and Crommie,  Michael F.},
  year = {2021},
  month = aug,
  pages = {1154–1161}
}

@article{Chen2020,
  title = {Strong correlations and orbital texture in single-layer {1T-TaSe$_2$}},
  volume = {16},
  ISSN = {1745-2481},
  url = {http://dx.doi.org/10.1038/s41567-019-0744-9},
  DOI = {10.1038/s41567-019-0744-9},
  number = {2},
  journal = {Nature Physics},
  publisher = {Springer Science and Business Media LLC},
  author = {Chen,  Yi and Ruan,  Wei and Wu,  Meng and Tang,  Shujie and Ryu,  Hyejin and Tsai,  Hsin-Zon and Lee,  Ryan L. and Kahn,  Salman and Liou,  Franklin and Jia,  Caihong and Albertini,  Oliver R. and Xiong,  Hongyu and Jia,  Tao and Liu,  Zhi and Sobota,  Jonathan A. and Liu,  Amy Y. and Moore,  Joel E. and Shen,  Zhi-Xun and Louie,  Steven G. and Mo,  Sung-Kwan and Crommie,  Michael F.},
  year = {2020},
  month = jan,
  pages = {218–224}
}

@article{Yao2024,
  title = {Moir\'e magnetism in {CrBr$_3$} multilayers emerging from differential strain},
  volume = {15},
  ISSN = {2041-1723},
  url = {http://dx.doi.org/10.1038/s41467-024-54870-2},
  DOI = {10.1038/s41467-024-54870-2},
  number = {1},
  journal = {Nature Communications},
  publisher = {Springer Science and Business Media LLC},
  author = {Yao,  Fengrui and Rossi,  Dario and Gabrovski,  Ivo A. and Multian,  Volodymyr and Hua,  Nelson and Watanabe,  Kenji and Taniguchi,  Takashi and Gibertini,  Marco and Gutiérrez-Lezama,  Ignacio and Rademaker,  Louk and Morpurgo,  Alberto F.},
  year = {2024},
  month = nov 
}

@article{Akram2024,
  title = {Theory of Moiré Magnetism in Twisted Bilayer {$\alpha$-RuCl$_3$}},
  volume = {24},
  ISSN = {1530-6992},
  url = {http://dx.doi.org/10.1021/acs.nanolett.3c04084},
  DOI = {10.1021/acs.nanolett.3c04084},
  number = {3},
  journal = {Nano Letters},
  publisher = {American Chemical Society (ACS)},
  author = {Akram,  Muhammad and Kapeghian,  Jesse and Das,  Jyotirish and Valentí,  Roser and Botana,  Antia S. and Erten,  Onur},
  year = {2024},
  month = jan,
  pages = {890–896}
}

@article{PhysRevLett.133.246703,
  title = {Stacking-Engineered Ferroelectricity and Multiferroic Order in van der {Waals} Magnets},
  author = {Bennett, Daniel and Mart\'{\i}nez-Carracedo, Gabriel and He, Xu and Ferrer, Jaime and Ghosez, Philippe and Comin, Riccardo and Kaxiras, Efthimios},
  journal = {Phys. Rev. Lett.},
  volume = {133},
  issue = {24},
  pages = {246703},
  numpages = {7},
  year = {2024},
  month = {Dec},
  publisher = {American Physical Society},
  doi = {10.1103/PhysRevLett.133.246703},
  url = {https://link.aps.org/doi/10.1103/PhysRevLett.133.246703}
}

@article{Yao2023,
  title = {Multiple antiferromagnetic phases and magnetic anisotropy in exfoliated {CrBr$_3$} multilayers},
  volume = {14},
  ISSN = {2041-1723},
  url = {http://dx.doi.org/10.1038/s41467-023-40723-x},
  DOI = {10.1038/s41467-023-40723-x},
  number = {1},
  journal = {Nature Communications},
  publisher = {Springer Science and Business Media LLC},
  author = {Yao,  Fengrui and Multian,  Volodymyr and Wang,  Zhe and Ubrig,  Nicolas and Teyssier,  Jérémie and Wu,  Fan and Giannini,  Enrico and Gibertini,  Marco and Gutiérrez-Lezama,  Ignacio and Morpurgo,  Alberto F.},
  year = {2023},
  month = aug 
}

@article{Fumega2023,
  title = {Moir\'e-driven multiferroic order in twisted {CrCl$_3$,  CrBr$_3$ and CrI$_3$} bilayers},
  volume = {10},
  ISSN = {2053-1583},
  url = {http://dx.doi.org/10.1088/2053-1583/acc671},
  DOI = {10.1088/2053-1583/acc671},
  number = {2},
  journal = {2D Materials},
  publisher = {IOP Publishing},
  author = {Fumega,  Adolfo O and Lado,  Jose L},
  year = {2023},
  month = apr,
  pages = {025026}
}

@article{Anto2024,
  title = {Electric Field Control Of Moiré Skyrmion Phases in Twisted Multiferroic {NiI$_2$} Bilayers},
  volume = {24},
  ISSN = {1530-6992},
  url = {http://dx.doi.org/10.1021/acs.nanolett.4c04582},
  DOI = {10.1021/acs.nanolett.4c04582},
  number = {49},
  journal = {Nano Letters},
  publisher = {American Chemical Society (ACS)},
  author = {Antão,  Tiago V. C. and Lado,  Jose L. and Fumega,  Adolfo O.},
  year = {2024},
  month = nov,
  pages = {15767–15773}
}

@article{Song2021,
  title = {Direct visualization of magnetic domains and moiré magnetism in twisted {2D} magnets},
  volume = {374},
  ISSN = {1095-9203},
  url = {http://dx.doi.org/10.1126/science.abj7478},
  DOI = {10.1126/science.abj7478},
  number = {6571},
  journal = {Science},
  publisher = {American Association for the Advancement of Science (AAAS)},
  author = {Song,  Tiancheng and Sun,  Qi-Chao and Anderson,  Eric and Wang,  Chong and Qian,  Jimin and Taniguchi,  Takashi and Watanabe,  Kenji and McGuire,  Michael A. and St\"{o}hr,  Rainer and Xiao,  Di and Cao,  Ting and Wrachtrup,  J\"{o}rg and Xu,  Xiaodong},
  year = {2021},
  month = nov,
  pages = {1140–1144}
}

@article{Chen_2022,
doi = {10.1088/1361-648X/ac9812},
url = {https://doi.org/10.1088/1361-648X/ac9812},
year = {2022},
month = {oct},
publisher = {IOP Publishing},
volume = {34},
number = {48},
pages = {485805},
author = {Chen, Guangze and Rösner, Malte and Lado, Jose L},
title = {Controlling magnetic frustration in {1T-TaS$_2$} via Coulomb engineered long-range interactions},
journal = {Journal of Physics: Condensed Matter}
}

@article{PhysRevLett.122.106404,
  title = {Origin of the Insulating Phase and First-Order Metal-Insulator Transition in {$1T\text{\ensuremath{-}}{\mathrm{TaS}}_{2}$}},
  author = {Lee, Sung-Hoon and Goh, Jung Suk and Cho, Doohee},
  journal = {Phys. Rev. Lett.},
  volume = {122},
  issue = {10},
  pages = {106404},
  numpages = {6},
  year = {2019},
  month = {Mar},
  publisher = {American Physical Society},
  doi = {10.1103/PhysRevLett.122.106404},
  url = {https://link.aps.org/doi/10.1103/PhysRevLett.122.106404}
}

@article{10.1063/1.4805003,
    author = {Liu, Y. and Ang, R. and Lu, W. J. and Song, W. H. and Li, L. J. and Sun, Y. P.},
    title = {Superconductivity induced by {Se}-doping in layered charge-density-wave system {1$T$-TaS$_{2-x}$Se$_x$}},
    journal = {Applied Physics Letters},
    volume = {102},
    number = {19},
    pages = {192602},
    year = {2013},
    month = {05},
    issn = {0003-6951},
    doi = {10.1063/1.4805003},
    url = {https://doi.org/10.1063/1.4805003}
}

@article{Sutter2019,
author={Sutter, Peter
and Ibragimova, Rina
and Komsa, Hannu-Pekka
and Parkinson, Bruce A.
and Sutter, Eli},
title={Self-organized twist-heterostructures via aligned van der {Waals} epitaxy and solid-state transformations},
journal={Nature Communications},
year={2019},
month={Dec},
day={04},
volume={10},
number={1},
pages={5528},
issn={2041-1723},
doi={10.1038/s41467-019-13488-5},
url={https://doi.org/10.1038/s41467-019-13488-5}
}

@article{PhysRevB.108.075416,
  title = {Electronic properties of twisted {$\mathrm{Sb}/\mathrm{W}{\mathrm{Te}}_{2}$ van der Waals} heterostructure with controllable band gap, band alignment, and spin splitting},
  author = {Wang, Fanfan and Yuan, Jun and Zhang, Zhufeng and Ding, Xunlei and Gu, Chenjie and Yan, Shubin and Sun, Jinhua and Jiang, Tao and Wu, Yifeng and Zhou, Jun},
  journal = {Phys. Rev. B},
  volume = {108},
  issue = {7},
  pages = {075416},
  numpages = {11},
  year = {2023},
  month = {Aug},
  publisher = {American Physical Society},
  doi = {10.1103/PhysRevB.108.075416},
  url = {https://link.aps.org/doi/10.1103/PhysRevB.108.075416}
}

@article{MARTINI2023106,
title = {Twisted cuprate van der {Waals} heterostructures with controlled {Josephson} coupling},
journal = {Materials Today},
volume = {67},
pages = {106-112},
year = {2023},
issn = {1369-7021},
doi = {https://doi.org/10.1016/j.mattod.2023.06.007},
url = {https://www.sciencedirect.com/science/article/pii/S1369702123001980},
author = {Mickey Martini and Yejin Lee and Tommaso Confalone and Sanaz Shokri and Christian N. Saggau and Daniel Wolf and Genda Gu and Kenji Watanabe and Takashi Taniguchi and Domenico Montemurro and Valerii M. Vinokur and Kornelius Nielsch and Nicola Poccia}
}

@article{PhysRevB.97.045133,
  title = {Charge density wave states in tantalum dichalcogenides},
  author = {Miller, David C. and Mahanti, Subhendra D. and Duxbury, Phillip M.},
  journal = {Phys. Rev. B},
  volume = {97},
  issue = {4},
  pages = {045133},
  numpages = {10},
  year = {2018},
  month = {Jan},
  publisher = {American Physical Society},
  doi = {10.1103/PhysRevB.97.045133},
  url = {https://link.aps.org/doi/10.1103/PhysRevB.97.045133}
}

@article{doi:10.1073/pnas.2520703123,
author = {Yonghao Liu  and Yuan Zheng  and Kun Yang  and Wenhao Zhang  and Zongxiu Wu  and Jingjing Gao  and Xuan Luo  and Yuping Sun  and Jin Zhang  and Yi Yin },
title = {Modulation of electronic structure via dual moir\'e patterns in twisted {1$T$-TaSe$_2$}},
journal = {Proceedings of the National Academy of Sciences},
volume = {123},
number = {11},
pages = {e2520703123},
year = {2026},
doi = {10.1073/pnas.2520703123},
URL = {https://www.pnas.org/doi/abs/10.1073/pnas.2520703123},
eprint = {https://www.pnas.org/doi/pdf/10.1073/pnas.2520703123}}

@article{LI20192302,
title = {Electronic structures of twist-stacked {1T-TaS$_2$} bilayers},
journal = {Physics Letters A},
volume = {383},
number = {19},
pages = {2302-2308},
year = {2019},
issn = {0375-9601},
doi = {https://doi.org/10.1016/j.physleta.2019.04.043},
url = {https://www.sciencedirect.com/science/article/pii/S0375960119303615},
author = {Yunting Li and Huaping Xiao and Pan Zhou and Juexian Cao}
}

@article{D0NR05148A,
author ="Li, Hui and Liu, Pan and Liu, Qi and Luo, Ruichun and Guo, Chenguang and Wang, Ziqian and Guan, Pengfei and Florencio Aleman, Christopher and Huang, Fuqiang and Chen, Mingwei",
title  ="Twisted {1T TaS$_2$} bilayers by lithiation exfoliation",
journal  ="Nanoscale",
year  ="2020",
volume  ="12",
issue  ="35",
pages  ="18031-18038",
publisher  ="The Royal Society of Chemistry",
doi  ="10.1039/D0NR05148A",
url  ="http://dx.doi.org/10.1039/D0NR05148A"}

@article{doi:10.1021/acs.chemrev.3c00627,
author = {Sun, Xueqian and Suriyage, Manuka and Khan, Ahmed Raza and Gao, Mingyuan and Zhao, Jie and Liu, Boqing and Hasan, Md Mehedi and Rahman, Sharidya and Chen, Ruo-si and Lam, Ping Koy and Lu, Yuerui},
title = {Twisted van der {Waals} Quantum Materials: Fundamentals, Tunability, and Applications},
journal = {Chemical Reviews},
volume = {124},
number = {4},
pages = {1992-2079},
year = {2024},
doi = {10.1021/acs.chemrev.3c00627},
    note ={PMID: 38335114},
URL = {https://doi.org/10.1021/acs.chemrev.3c00627},
eprint = {https://doi.org/10.1021/acs.chemrev.3c00627}
}

@article{Dalal2025,
author={Dalal, Amir
and Ruhman, Jonathan
and Venderbos, J{\"o}rn W. F.},
title={Flat band physics in the charge-density wave state of {1T-TaS$_2$} and {1T-TaSe$_2$}},
journal={npj Quantum Materials},
year={2025},
month={Mar},
day={20},
volume={10},
number={1},
pages={31},
issn={2397-4648},
doi={10.1038/s41535-025-00747-6},
url={https://doi.org/10.1038/s41535-025-00747-6}
}

@article{shah_progress_2024,
	title = {Progress and prospects of {Moiré} superlattices in twisted {TMD} heterostructures},
	volume = {17},
	issn = {1998-0000},
	url = {https://doi.org/10.1007/s12274-024-6936-3},
	doi = {10.1007/s12274-024-6936-3},
	number = {11},
	urldate = {2026-06-23},
	journal = {Nano Research},
	author = {Shah, Syed Jamal and Chen, Junying and Xie, Xing and Oyang, Xinyu and Ouyang, Fangping and Liu, Zongwen and Wang, Jian-Tao and He, Jun and Liu, Yanping},
	month = nov,
	year = {2024},
	pages = {10134--10161},
}

@article{huang_angle-dependent_2024,
	title = {Angle-{Dependent} {Oscillatory} {Interlayer} {Coupling} in {Twisted} {MoS2}/{TaSe2} {Heterostructure}},
	volume = {128},
	issn = {1932-7447},
	url = {https://doi.org/10.1021/acs.jpcc.3c08154},
	doi = {10.1021/acs.jpcc.3c08154},
	number = {8},
	urldate = {2026-06-23},
	journal = {The Journal of Physical Chemistry C},
	publisher = {American Chemical Society},
	author = {Huang, Xixi and Luo, Siwei and Guo, Gencai and Zhang, Jibo and Ren, Long and Chen, Qiong and Zhong, Jianxin and Qi, Xiang},
	month = feb,
	year = {2024},
	pages = {3393--3399},
}

@article{dai_layer_2023,
	title = {Layer {Sliding} {And} {Twisting} {Induced} {Electronic} {Transitions} {In} {Correlated} {Magnetic} 1t-{Nbse2} {Bilayers}},
	volume = {33},
	copyright = {© 2023 Wiley-VCH GmbH},
	issn = {1616-3028},
	url = {https://onlinelibrary.wiley.com/doi/abs/10.1002/adfm.202302989},
	doi = {10.1002/adfm.202302989},
	number = {38},
	urldate = {2026-06-23},
	journal = {Advanced Functional Materials},
	author = {Dai, Jiaqi and Qiao, Jingsi and Wang, Cong and Zhou, Linwei and Wu, Xu and Liu, Liwei and Song, Xuan and Pang, Fei and Cheng, Zhihai and Kong, Xianghua and Wang, Yeliang and Ji, Wei},
	year = {2023},
	pages = {2302989},
}

@article{hu_engineering_2022,
	title = {Engineering chiral topological superconductivity in twisted {Ising} superconductors},
	volume = {106},
	issn = {2469-9950, 2469-9969},
	url = {https://link.aps.org/doi/10.1103/PhysRevB.106.125136},
	doi = {10.1103/PhysRevB.106.125136},
	number = {12},
	urldate = {2026-06-24},
	journal = {Physical Review B},
	author = {Hu, Xiaodong and Ran, Ying},
	month = sep,
	year = {2022},
	pages = {125136},
}

@article{kovalska_photocatalytic_2021,
	title = {Photocatalytic activity of twist-angle stacked {2D} {TaS2}},
	volume = {5},
	copyright = {2021 The Author(s)},
	issn = {2397-7132},
	url = {https://www.nature.com/articles/s41699-021-00247-8},
	doi = {10.1038/s41699-021-00247-8},
	number = {1},
	urldate = {2026-06-24},
	journal = {npj 2D Materials and Applications},
	publisher = {Nature Publishing Group},
	author = {Kovalska, Evgeniya and Roy, Pradip Kumar and Antonatos, Nikolas and Mazanek, Vlastimil and Vesely, Martin and Wu, Bing and Sofer, Zdenek},
	month = jul,
	year = {2021},
	pages = {68},
}

@article{yuan_probing_2023,
	title = {Probing the {Twist}-{Controlled} {Interlayer} {Coupling} in {Artificially} {Stacked} {Transition} {Metal} {Dichalcogenide} {Bilayers} by {Second}-{Harmonic} {Generation}},
	volume = {17},
	issn = {1936-0851},
	url = {https://doi.org/10.1021/acsnano.3c03795},
	doi = {10.1021/acsnano.3c03795},
	number = {18},
	urldate = {2026-06-24},
	journal = {ACS Nano},
	publisher = {American Chemical Society},
	author = {Yuan, Yuanjian and Liu, Peng and Wu, Hongjian and Chen, Haitao and Zheng, Weihao and Peng, Gang and Zhu, Zhihong and Zhu, Mengjian and Dai, Jiayu and Qin, Shiqiao and Novoselov, Kostya S.},
	month = sep,
	year = {2023},
	pages = {17897--17907},
}

@article{weston_atomic_2020,
	title = {Atomic reconstruction in twisted bilayers of transition metal dichalcogenides},
	volume = {15},
	copyright = {2020 The Author(s), under exclusive licence to Springer Nature Limited},
	issn = {1748-3395},
	url = {https://www.nature.com/articles/s41565-020-0682-9},
	doi = {10.1038/s41565-020-0682-9},
	number = {7},
	urldate = {2026-06-24},
	journal = {Nature Nanotechnology},
	publisher = {Nature Publishing Group},
	author = {Weston, Astrid and Zou, Yichao and Enaldiev, Vladimir and Summerfield, Alex and Clark, Nicholas and Zólyomi, Viktor and Graham, Abigail and Yelgel, Celal and Magorrian, Samuel and Zhou, Mingwei and Zultak, Johanna and Hopkinson, David and Barinov, Alexei and Bointon, Thomas H. and Kretinin, Andrey and Wilson, Neil R. and Beton, Peter H. and Fal’ko, Vladimir I. and Haigh, Sarah J. and Gorbachev, Roman},
	month = jul,
	year = {2020},
	pages = {592--597},
}

@article{PhysRevLett.102.256802,
  title = {Theory of Single-Spin Inelastic Tunneling Spectroscopy},
  author = {Fern\'andez-Rossier, J.},
  journal = {Phys. Rev. Lett.},
  volume = {102},
  issue = {25},
  pages = {256802},
  numpages = {4},
  year = {2009},
  month = {Jun},
  publisher = {American Physical Society},
  doi = {10.1103/PhysRevLett.102.256802},
  url = {https://link.aps.org/doi/10.1103/PhysRevLett.102.256802}
}

\end{document}